\documentclass[12pt]{article}
\usepackage{graphics,amssymb}

\newcommand{\bea}{\begin{eqnarray}}
\newcommand{\eea}{\end{eqnarray}}
\newcommand{\be}{\begin{equation}}
\newcommand{\ee}{\end{equation}}

\def\be{\begin{eqnarray}}
\def\ee{\end{eqnarray}}
\def\bd{\begin{displaymath}}
\def\ed{\end{displaymath}}

\def\NP{Nucl. Phys. }
\def\PR{Phys. Rev. }
\def\PRL{Phys. Rev. Lett. }
\def\PL{Phys. Lett. }
\def\jpg{J. Phys. G: Nucl. Part. Phys. }

\begin{document}

\title{Correction to Relativistic Mean Field binding energy and $N_pN_n$ scheme}

\author{Madhubrata Bhattacharya and
 G. Gangopadhyay\\
Department of Physics, University of Calcutta\\
92, Acharya Prafulla Chandra Road, Kolkata-700 009, India\\
email: ggphy@caluniv.ac.in}
\date{}

\maketitle

\begin{abstract}
The differences between the experimental and Relativistic Mean Field 
binding energies have been calculated for a 
large number of even-even nuclei from $A=50$ to 220. Excluding certain
mass regions, the differences, after suitable corrections for particular
isotope chains, are found to be proportional to the Casten factor $P$,
chosen as a measure of n-p interaction strength in a nucleus.
Results for even-$Z$ odd-$N$ nuclei are also seen to follow the same relation,
if the odd-even mass difference is taken into account following the 
semiempirical formula. This indicates that the n-p interaction is the major
contributor to the difference between the calculated and the experimental
binding energies.
\end{abstract}

PACS 21.10.Dr,21.60.Jz
 
Keywords: Binding Energy, Relativistic Mean Field, n-p interaction

It is well known that simplified parametrization of various nuclear 
quantities are obtained as functions of 
$N_pN_n$, the product of effective number of valance particles (or holes)
\cite{Casten1}. 
Essentially this simple product is seen to represent integrated n-p 
interaction strength and to bear 
smooth relationships with the observables.
The correlations beyond mean field results are due principally
to residual two body interaction. In a mean field calculation, the residual 
interaction between similar nucleons is  described by the pairing force. 
However, the calculations usually ignore the residual n-p interaction. 
For a chain of isotopes, the difference between the experimental and the
calculated binding energies may be a measure of the integrated strength of n-p 
interaction in a particular nucleus and vary smoothly with certain simple 
functions of $N_p$ and $N_n$.

Various quantities such as deformation and B(E2) values
\cite{Casten3,Foy,Zhao}, rotational moments of inertia in low spin states in 
the rare earth region\cite{Saha}, ground band energy systematics \cite{Saha1}, 
core cluster decomposition in the rare earth region\cite{Buck}, and properties 
of excited states \cite{Casten2,yoon} have been found to follow certain simple 
trends when expressed as a function of  the product of $N_p$ and $N_n$ or 
certain simple functions of the above two quantities. In the present work, we 
attempt to show that binding energy corrections to Relativistic Mean Field (RMF)
calculations can also be expressed in a similar fashion.

However, not all the difference between the experimental and the theoretical 
binding energies can be ascribed to the effect of n-p interaction. To extract 
this effect, we have selected
the isotope for each Z with magic neutron number 
{\em i.e.} isotopes with no valence n-p pairs. In these nuclei, we expect 
the effect of n-p interaction to be small and the difference between
the experimental and calculated binding energies to be due to
all the other effects combined. The difference between theory and experiment 
in the change in the binding energy from the isotope with $N_n=0$ for a 
particular Z is
taken as a measure of the contribution of $N_pN_n$ interaction and expressed as
$\Delta_{\nu\pi}$. Thus we write
\be \Delta_{\nu\pi}(Z,N)=A(B_{th}(Z,N)-B_{ex}(Z,N)+B_{corr}(Z))\ee
where, $B_{th}$ and $B_{ex}$ are respectively the theoretically
calculated and experimentally measured binding energies per nucleon and, 
$A=Z+N$, the mass number. We have defined  
$B_{corr}(Z)=B_{ex}(Z,N_0)-B_{th}(Z,N_0)$, 
$N_0$ being a magic number. Depending on the 
neutron core, the quantity $B_{corr}(Z)$ may have more than one value. For 
example, for Cd isotopes with $N\ge 66$, one has to use the experimental and 
theoretical binding energy values for 
the isotope with $N=82$ while for the lighter isotopes, one uses the values
for $N=50$. Obviously $\Delta_{\nu\pi}(Z,N)$ vanishes for magic $N$. The 
experimental binding energy values are from Ref. \cite{mass}.

There exist different variations of the Lagrangian density as
well as a number of different parametrization in RMF.  The Lagrangian
density FSU Gold\cite{prl}, which involves
self-coupling of the vector-isoscalar meson as well as coupling between the
vector-isoscalar meson and the vector-isovector meson,
was earlier employed in our
study of proton radioactivity\cite{plb}, 
alpha radioactivity in heavy and superheavy nuclei\cite{282,plb2}, and cluster 
radioactivity\cite{cluster}. In Ref \cite{plb2}, spectroscopic factors and
$\Delta_{\nu\pi}$ values in actinides were seen to
follow a certain pattern.
In that region the only appropriate major doubly closed shell nucleus is
$^{208}$Pb and it was necessary to employ subshell closures. In the
present work we look for a more robust systematics in $\Delta_{\nu\pi}$, 
valid in a large
mass region and dependent only on the known major shells.
The FSU Gold Lagrangian density seems very appropriate
for a large mass region {\em viz}. medium mass to superheavy nuclei. 
We have solved the equations in co-ordinate space. The strength of the zero 
range pairing force is taken as 300 MeV-fm for both
protons and neutrons. We have also checked our conclusions using the density
NL3\cite{NL3} which gives very similar results. Unless otherwise mentioned,
the results refer to the calculations with FSU Gold.

\begin{figure*}
\leftskip -0.2cm
\resizebox{!}{!}{ \includegraphics{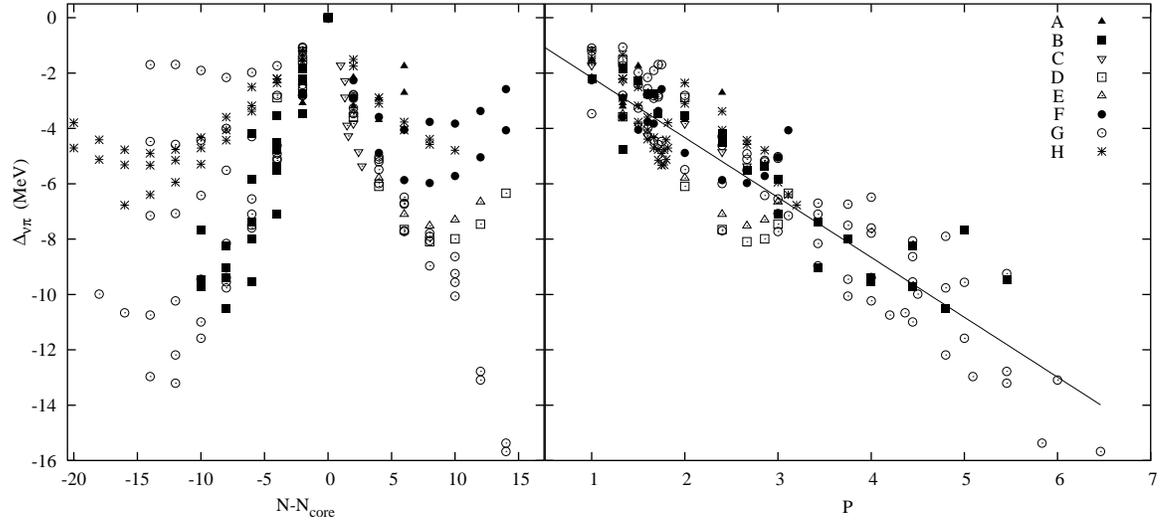}}
\caption{$\Delta_{\nu\pi}$ as a function of N-N$_{core}$ (left hand plot) and 
$P=N_pN_n/(N_p+N_n)$ (righthand plot).
Symbols used for nuclei in different mass regions are indicated in Table I.
\label{bediffgold}}
\end{figure*}

In Fig. \ref{bediffgold}, we plot the results of a large number of even-even
nuclei, lying between mass 50 and mass 220 as shown in Table \ref{list}. The 
results have been plotted only for the nuclei whose experimental binding 
energies are available. Certain isotope chains, {\em e.g.} the chains of 
isotopes for $Z = 64-70$ and $88\le Z\le92$, do not follow the pattern
that we have observed in the nuclei of Table \ref{list} and have been discussed 
later. Values of $\Delta_{\nu\pi}$ could not be calculated
for certain nuclei as experimental binding energies for the isotopes with 
$N_n=0$ are not available and have been treated separately.
In the left hand plot of Fig. \ref{bediffgold}, we have plotted the quantity 
$\Delta_{\nu\pi}$ as a function of number of $N-N_{\rm core}$, where $N$ is the 
number of neutrons and $N_{\rm core}$ is the nearest closed neutron shell. 
It is difficult to see a pattern
for the different mass regions, or even, within a mass region. However, we 
find that the points lie very close
to a straight line if plotted as a function of the Casten factor,
$P=N_pN_n/(N_p+N_n)$ which has been widely used as a measure of 
the integrated n-p interaction strength. In fact the quantity may be expressed 
as simply proportional to $P$. One can fit a straight line 
\be \Delta_{\nu\pi}=aP\ee
 with $a=-2.148\pm 0.029$
with rms deviation 1.15 MeV. The fitting does not include the values
for nuclei with $P=0$ which are defined to be zero.  The fitted line has been 
shown in 
the right panel of Fig. \ref{bediffgold}. In a few cases, to improve the 
results, certain shell  closures, which are not apparent, are chosen. 
For example, in lower $Z$ nuclei among those represented by `C', proton shell 
closure is 38, and not 20 or 28. However, in most situations, the choice of the 
magic number is self-evident.

The theoretical values may be corrected using the fitted straight line in eqn. 
(2) enormously improving the agreement between the calculated and experimental
binding energy values. It is worth noting that that the present 
mean field calculation does not take deformation into account and is expected 
to underpredict the binding energy far away from the closed shell. However, 
with this correction from eqn. (2), it is possible to obtain
an agreement comparable to or even better than the values calculated using a 
deformed mean field approach.

It is possible to extend our calculation to situation where the experimental 
binding energy for the isotope with magic neutron number is not known.
The nuclei, with the
proton and neutron magic numbers chosen to calculate $N_p$ and $N_n$ 
given in parentheses, $^{112-120}$Pd(50,82),  $^{110-116}$Te(50,50),
 $^{112-118}$Xe(50,50), $^{114-120}$Ba(50,50) have been studied. We also 
include all the nuclei
with $N\ge 106$ and $Z=70-78$, all with the same magic core (82,126), whose 
experimental 
binding energies are known
{\em i.e.} $^{176,178}$Yb, $^{178-184}$Hf, $^{180-190}$W,
 $^{182-196}$Os, and $^{184-200}$Pt.

\begin{figure}
\resizebox{7.4cm}{!}{\includegraphics{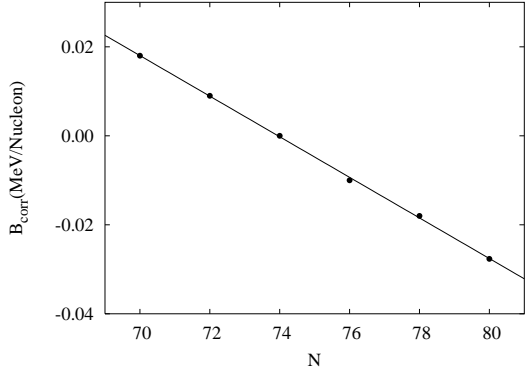}}
\caption{$B_{corr}$ values for $Z=70-80$. See text for details.
\label{xplot}}
\end{figure}

\begin{figure}
\resizebox{7.4cm}{!}{\includegraphics{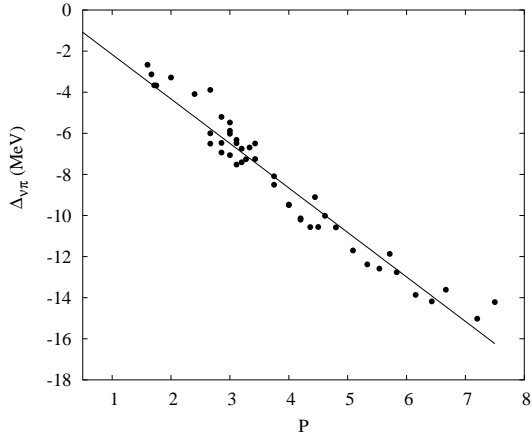}}
\caption{$\Delta_{\nu\pi}$ as a function of $P$ for the isotopes
$Z=46,~N\ge66$; $Z=52-56,~N\le64$; and $Z=70-78,~N\ge 106$ as described in the 
text.\label{pre}} 
\end{figure}

The $B_{corr}$ values for the above chains may be estimated in two ways. It 
may be taken from a different shell closure where the experimental data is 
available. For example,the binding energies for Te, Xe and Ba nuclei with $N=50$
are obviously not available as they lie beyond the proton drip line. However, 
the $B_{corr}$ 
values for these nuclei with $N=82$ have already been calculated in the present 
work and we use the same values for  the nuclei mentioned above. 
In Pd nuclei, the value obtained from $N=50$ cannot be 
used  for the $N=82$ shell closure and is actually calculated in the following 
approach. 
In nuclei with 
$Z=70-78$, the experimental binding energy is not available for $N=126$. 
The binding energy for $^{152}$Yb is known, but the Yb isotopes in its vicinity
do not share the simple trend of eqn. (2). 
In  nuclei with $Z=46$ and $74-78$ we have estimated $B_{corr}$  from the 
differences between the theoretical and experimental binding energies in 
isotopes with $N_n\neq0$ by using eqn. (2) with the fitted value for $a$. 
For $Z=70$ and 72, the number of available $\Delta_{\nu\pi}$ values are 
rather small to extract $B_{corr}$ meaningfully. However, we find that the 
values of $B_{corr}$ obtained for 
$Z=74-78$ along with that obtained from the theoretical and experimental 
binding energy values of $^{206}_{80}$Hg lie on a straight line. We have 
obtained the values for $Z=70$ and $Z=72$ from the fitted line. The values of 
$B_{corr}$ used for $Z=70-80$ have been shown in Fig. \ref{xplot}.
The $\Delta_{\nu\pi}$
values for the above nuclei have been plotted against $P$ in Fig. \ref{pre}. 
Once again, one can see the excellent agreement between the extracted 
values of $\Delta_{\nu\pi}$ and the straight line of eqn. (2) also shown 
in the figure, plotted with the previously fitted value of $a$.

\begin{figure*}
\resizebox{7.4cm}{!}{ \includegraphics{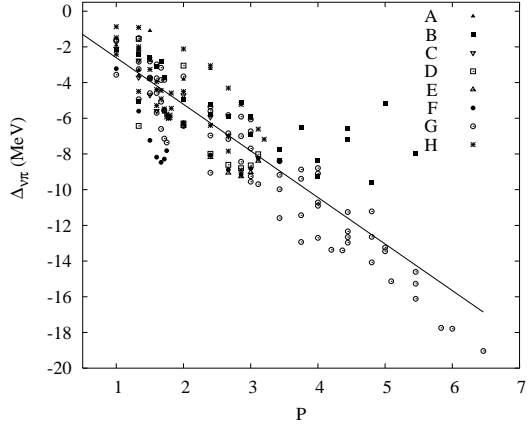}}
\caption{$\Delta_{\nu\pi}$ as a function of $P$ for the nuclei of Fig.
\ref{bediffgold} for the density NL3.
\label{bediffNL3}}
\end{figure*}

To check whether this remarkable correlation is a property of the particular
Lagrangian density alone, we have chosen another Lagrangian density, NL3 and studied 
the nuclei for which results have been plotted in Fig. \ref{bediffgold}. The 
results, shown in Fig. \ref{bediffNL3}, show a very similar trend though with 
slightly different slope ($a=-2.609\pm 0.044$) and a slightly higher 
 rms deviation of 1.68 MeV.
We have also compared our results with those of a deformed 
RMF calculation by Lalazissis {\em et al}
\cite{lalazissis} for Nd and Sm isotopes. We find that the agreement 
in binding energies and two nucleon separation 
energies using the present approach is comparable to or better than that 
observed in the deformed calculation.

\begin{figure}
\resizebox{7.4cm}{!}{\includegraphics{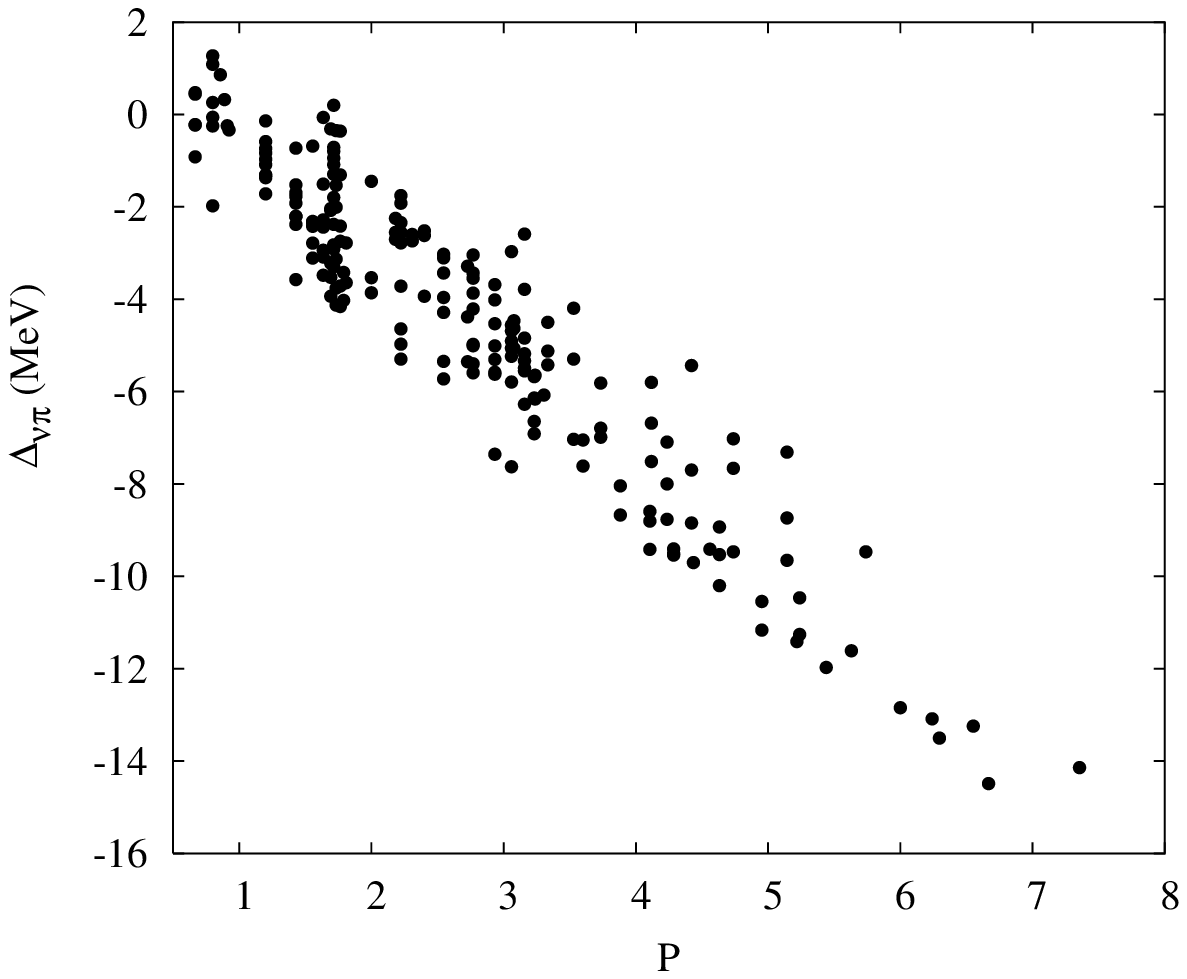}}
\caption{$\Delta_{\nu\pi}$ as a function of $P$ for odd-even isotopes
as described in the text.\label{odd}}
\end{figure}

\begin{figure}
\resizebox{7.4cm}{!}{\includegraphics{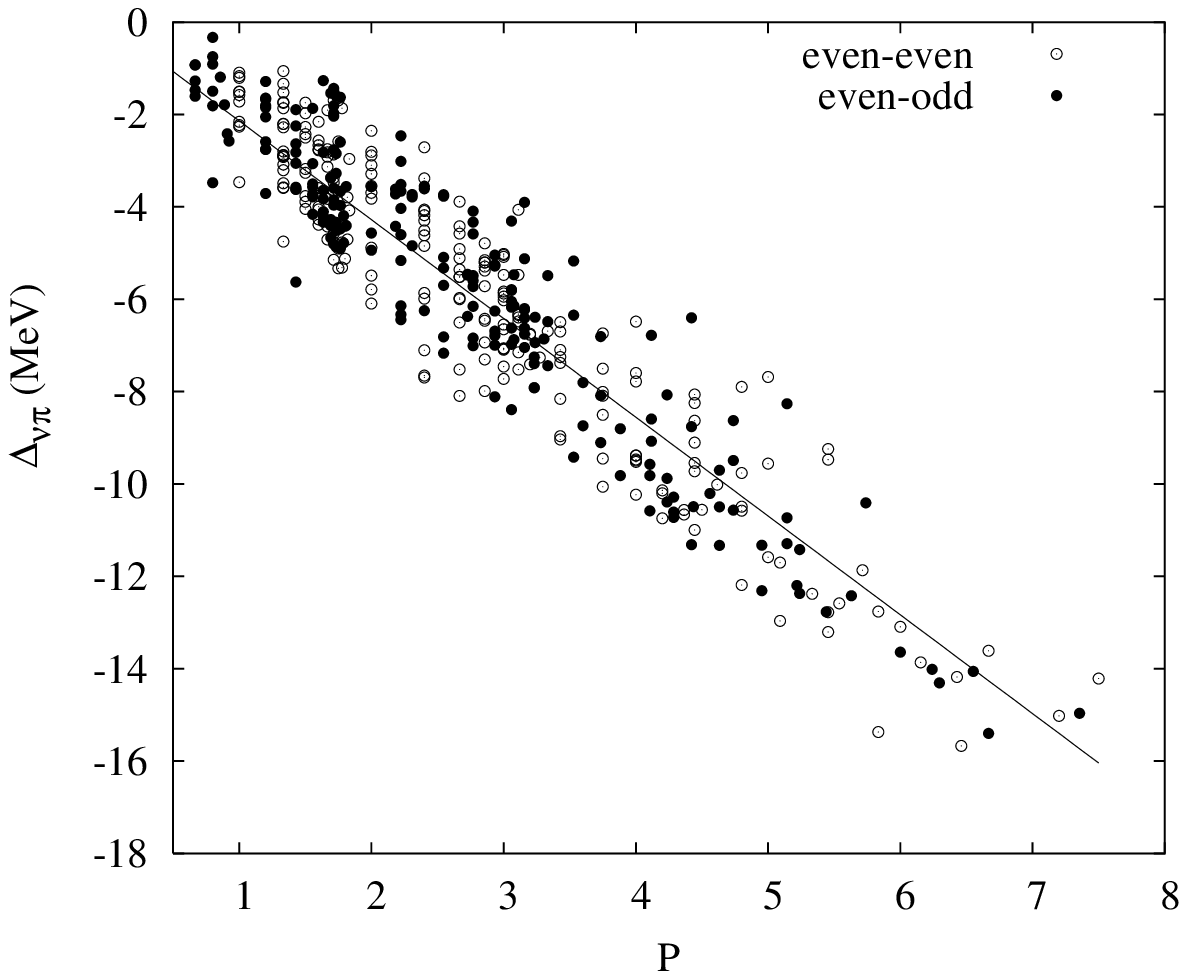}}
\caption{$\Delta_{\nu\pi}$ as a function of $P$ for even-even and odd-even 
isotopes as described in the text.\label{all}}
\end{figure}

The excellent results for even-even isotopes have prompted us to study
even-$Z$ odd-$N$ isotopes. This has the added advantage that the $B_{corr}(Z)$
values are already known from the study of the even-even chains. We have 
studied the odd N even Z isotopes within the ranges given in Table I. 
Additionally, we calculate $\Delta_{\nu\pi}$ values for the ranges
of isotopes discussed earlier where the binding energy values for the  
isotope with magic neutron number are not known and  $B_{corr}(Z)$
values have been estimated. In no case we have  modified the  $B_{corr}(Z)$
values for odd isotopes. In our calculation, we neglect the fact that, 
the unpaired neutron  actually occupies a particular single particle state,
and breaks the symmetry. However, it is known that the effect of this 
correction to the binding energy is small.
The results, plotted in Fig. \ref{odd}, again show a 
similar trend for even-odd isotopes. Keeping the odd-even mass difference 
term in the semiempirical mass formula in mind, we try to fit 
the results using a simple function of the form $aP+d/A$, where A is the 
mass number 
of the isotope. A least square fitting procedure gives the values as 
$a=-2.129\pm0.042$ and $d=145.7\pm14.3$ with a standard deviation of 1.09 MeV
for 209 nuclei. There are two points of interest here. The coefficients for the 
Casten factor $P$ for even-even and even-odd isotopes are identical within 
errors. Secondly, the value for $d$ is nearly the same as the corresponding 
coefficient in semi-empirical mass formula, i.e. 140 MeV. In Fig. \ref{all}, 
the results for all the isotopes described so far, except the ones with $P=0$, 
have been plotted. The results for the even-odd isotopes have been shifted by 
the amount $-145.7/A$. A least square fit of the points using eqn. (2) leads to 
a value, $a=-2.139\pm0.017$, with rms deviation of 1.09 MeV for 443
nuclei and have also been shown. Fig. \ref{all} clearly demonstrates that the 
$n-p$ interaction is the dominating factor in the correction to the RMF binding 
energy.

\begin{figure}
\resizebox{7.4cm}{!}{ \includegraphics{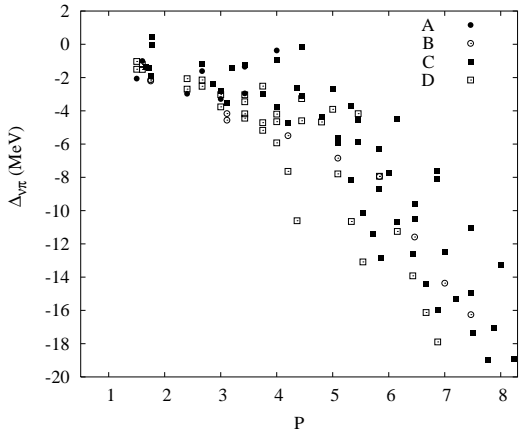}}
\caption{$\Delta_{\nu\pi}$ as function of $P$ for the nuclei as indicated
with the closed core given in parentheses.
{\bf A:} $Z=30,32, ~42\le N\le 50$ (38,50); {\bf B:} $Z=64,~ 78\le N\le98$
(50,82); {\bf C:} $Z=66-74,~ 82\le N\le104 $(82,82); {\bf D:} $88\le Z\le 92, ~114\le N \le148 $(82,126)
\label{bedifheavy}}
\end{figure}

Finally we would like to make a brief comment on the nuclei in various  mass 
regions not included in the above discussion, particularly the
rare earth nuclei $Z=64-74$, $N=78-104$ and actinide nuclei $Z=88-92$,
$N=114-148$.
The $\Delta_{\nu\pi}$ values for even-even nuclei in these regions follow a different trend as 
shown in Fig. \ref{bedifheavy}. First of all, the dispersion in the values is larger that
the case of lighter nuclei. More importantly, clearly there are two different
trends in the values with the points beyond $P=5$ showing a sharp downward 
tendency. 

Subshell closures, such as $Z=38$ or 64, often become important in the 
systematics of certain observables\cite{Casten1,jg}. As mentioned earlier, we 
also invoked a number of different of subshell closures in our work on 
systematics of spectroscopic factors\cite{plb2}. In the present work, we have 
already used $Z=38$ as a closure. We note that among the nuclei mentioned in 
the preceding paragraph, the subshell closure $Z=64$ brings the 
$\Delta_{\nu\pi}$ values for nuclei with $Z=66,68$,$82\le N\le 92$, very close 
to the straight line in Fig. \ref{all}. However, a more detailed analysis is
 required to bring out the role of subshell closures in the binding energy
corrections.

The differences between the experimental and the theoretically 
calculated binding energies in RMF approach have been 
calculated for a large number of even-even nuclei from $A=50$ to 220. 
As the n-p interaction is the major
contributor to the difference between the theoretical and the experimental
binding energies in RMF, we have taken the Casten factor $P$ as a measure of 
n-p interaction and found that excluding certain
mass regions, the differences, after suitable corrections for particular
isotope chains, are  proportional to  $P$.
Results for even-$Z$ odd-$N$ nuclei are also seen to follow the same relation,
if the odd-even mass difference 
is taken into account.

This work is carried out with financial assistance of the
Board of Research in Nuclear Sciences, Department of Atomic Energy (Sanction
No. 2005/37/7/BRNS).

\newpage
\begin{table}
\caption{Symbols used in Fig. 1 for nuclei in different mass regions and
the magic proton and neutron numbers used to calculate $N_p$ and $N_n$ for them.\label{list}}
\begin{tabular}{cllc}
Symbol & \multicolumn{1}{c}{$Z$-range} & $N$-range & Core($Z,N$)\\\hline
A & 22 - 24 & 26 - 34 & 20, 28\\
B & 26 - 36 & 30 - 40 & 38, 40\\
C & 34,36  & 42 - 50 & 38,50\\
D &  42 & 46 - 64 & 38, 50\\
E & 44 & 50 - 64 & 40, 50\\
F & 46 - 48 & 50 - 64& 50, 50\\
G & 48, 52 - 62 & 66 - 98 & 50, 82\\
H & 80, 84-86 & 106-136 & 82, 126\\
\hline
\end{tabular}
\end{table}

\pagebreak

 
\end{document}